\documentclass[journal,10pt]{elsarticle}
\pdfoutput=1
\biboptions{sort&compress,square}
\usepackage{graphicx}
\usepackage{lineno}
\newcommand{\fem}{f_{em}}
\newcommand{\vev}[1]{\left\langle#1\right\rangle}

\usepackage{color}
\usepackage{amssymb}

\begin{document}

\journal{Nuclear Instruments and Methods A}

\begin{frontmatter}

\title{Simplification of the DREAM collaboration's ``Q/S method'' 
in dual readout calorimetry analysis}

\author{Donald E. Groom}

\address{Lawrence Berkeley National Laboratory,
50R6008, Berkeley, CA 94720, USA}
\corref{Tel: 1 510 486 6788; fax: +1 510 486 4799}
\ead{degroom@lbl.gov}

\begin{abstract}
The DREAM collaboration has introduced the ``$Q/S$ Method'' for obtaining the energy 
estimator from simultaneous Cherenkov  and scintillator readouts of individual hadronic events. 
We show that the algorithm is equivalent to an elementary method.%
{\color{red} Corrections are indicated in red.}
\end{abstract}


\end{frontmatter}

\section{Introduction} 

The response of a hadronic calorimeter to an incident pion or jet of energy $E$ can be written as
\begin{equation}
\hbox{hadronic response} = E  [\,\fem+(1-\fem)   (h/e) ]
\label{eqn:hadresponse}
\end{equation} %
where a fraction $\fem$ is deposited in electromagnetic (EM) cascades, mostly initiated 
by $\pi^0$ decay gamma rays, and $h/e$ is the energy-independent ratio of detection 
efficiencies for the hadronic and EM energy deposits.\footnote{Whether one writes this
ratio as $e/h$, as is conventional, or $h/e$ is usually unimportant.  
This is not the case in Sec.~\ref{sec:discuss}, where we regard $h/e$ as a stochastic variable.}
(Here and elsewhere the energy $E$ is normalized to the electron response.)  In the case of a
dual readout calorimeter, in which a Cherenkov signal $Q$ and scintillator signal $S$ are 
read out for each event, Eq.~\ref{eqn:hadresponse} can be 
generalized:\cite{Wigmans_Perugia_04,DREAM05,groom07,RPP12}.
%{\color{blue}
\begin{eqnarray} 
Q &=&  E[ f_{em} + (1-f_{em})   (h/e|_Q)]  
\label{eqn:Q0}\\
S&=& E[  f_{em}+ (1- f_{em})   (h/e|_S) ]
\label{eqn:S0}
\end{eqnarray}
John Hauptman has suggested the less cumbersome notation $\eta_X \equiv (h/e|_X)$,
which we use in this paper:
\begin{eqnarray} 
Q &=&  E[ f_{em} + (1-f_{em}) \, \eta_Q] 
\label{eqn:Q}\\
S &=& E [f_{em}+ (1- f_{em})  \, \eta_S]
\label{eqn:S}
\end{eqnarray}
%}%
The EM fraction $\fem$ is a feature of the event, while the efficiency ratios 
$\eta_Q$ and $\eta_S$ are different for the two channels.  
Equations~\ref{eqn:Q} and~\ref{eqn:S} are the starting point
for any analysis of dual-readout  hadron calorimetry data.  

If $\fem$, $\eta_Q$, and $\eta_S$ are known exactly, and if there are no photoelectron 
or other statistical contributions, then $Q$ and $S$ are uniquely determined by the incident 
hadron energy $E$.  If, on the other hand, all of these quantities are subject to statistical 
fluctuations, then $E$ as determined from the equations must be regarded as the \it estimator\rm\
of the hadron (or jet) energy for a particular event.

These equations appear explicitly in Fig.~11 of the first DREAM paper at the Perugia
Conference on Calorimetry in High Energy Physics\cite{Wigmans_Perugia_04} and appear
either explicitly or implicitly in subsequent DREAM papers.  The most complete description 
of the DREAM analysis is given by Akchurin, et al.\cite{DREAM05} (henceforth Ak05),
and it is the basic reference for this paper. 
Several data reduction schemes are described, but the algorithm considered most basic is 
the fairly convoluted  ``Energy-independent $Q/S$ 
correction method.'' We show here that it can be obtained in a few lines from 
Eqns.~\ref{eqn:Q} and~\ref{eqn:S}.

\section{The energy-independent $Q/S$ correction method.} \label{QoverS}

The estimator  $E$, whose determination is the object of the analysis, can be 
eliminated by dividing Eq.~\ref{eqn:Q} by Eq.~\ref{eqn:S}, to obtain
\begin{equation}
\frac{Q}{S} = \frac{ f_{em} + (1-f_{em})   \eta_Q}{f_{em} + (1-f_{em})   \eta_S} \  . 
\end{equation}
This is Eq.~2 in Ak05, except that in that paper values of $h/e$ special to the DREAM 
experiment are inserted for $\eta_Q$ and $\eta_S$.
It can be solved for $\fem$. Although the result, 
\begin{equation}
\fem = \frac      {(Q/S)\eta_S-\eta_Q}     {(1-\eta_Q) -(Q/S)(1-   \eta_S)  } \ ,
\end{equation}
is not given in the paper, its availability is assumed in the rest of its
 discussion.

Leakage corrections are incorporated as part of the Method. They are obviously important, but here we assume they have already been made to $Q$ and $S$ as given in 
Eqns.~\ref{eqn:Q} and~\ref{eqn:S}.

The final estimator of the energy, called $S_{\rm final}$, is given by Ak05's Eq.~7:
\begin{equation}
S_{\rm final} = S_{\rm corr}\left[\frac{1+p_1/p_0}{1+\fem\,p_1/p_0}\right] \ ,
\end{equation}
where $p_1/p_0 = e/h-1$.  We identify $S_{\rm final}$ with the energy estimator $E$, and 
replace $S_{\rm corr}$ by $S$ because the leakage corrections hve already been made. 
From context, $e/h$ is $e/h|_S$. The equation then becomes
\begin{eqnarray}
E &= & S\,\frac{e/h|_S}{1 + \fem(e/h|_S -1)}\\
&=& \frac{S}{\eta_S + \fem(1-\eta_S)}\ ,
\end{eqnarray}\
which we recognize as just a rearrangement of Eq.~\ref{eqn:S}.

%} % end phantom
It remains to insert the expression for $\fem$ into this equation.
Simplification of the result is fairly tedious, but finally yields
\begin{equation}
E = S\,\left[ \frac{(1-\eta_Q) - (Q/S){\color{red}(1-}\eta_S{\color{red})}}{\eta_S - \eta_Q}\right]  \ .
\label{eqn:EviaQoverS}
\end{equation}

\section{Direct solution}

We can write the simultaneous equations \ref{eqn:Q} and \ref{eqn:S} as 
\begin{equation}
\left( \matrix{ Q &-(1- \eta_Q)\cr S& -(1- \eta_S) }\right)
 \left( \matrix{1/E \cr \fem }\right)  
= \left( \matrix{\eta_Q \cr \eta_S }\right) 
\label{eqn:matrixversion}
\end{equation}
with immediate solutions
\begin{eqnarray}
\nonumber E &=&S\,\left[\displaystyle \frac {(1- \eta_Q) - (Q/S){\color{red}(1-}\eta_S{\color{red})}  }  {\eta_S - \eta_{\color{red}Q}    } \right] \\
&\color{blue}=&\displaystyle\color{blue} \frac {S(1- \eta_Q) - Q(1-\eta_S)}{\eta_S -  \eta_Q}\\
\label{eqn:Elongform} 
\nonumber\fem &=& \frac      {(Q/S)\eta_S-\eta_Q}     {(1-\eta_{\color{red}Q}) -(Q/S)(1-   \eta_{\color{red}S})  }\\
\label{eqn:femSolution}
&\color{blue}=&\color{blue} \displaystyle\frac      {Q\eta_S-S\eta_Q}     {S(1-\eta_Q) -Q(1-   \eta_S)  }
\end{eqnarray}
for the \it estimators \rm  of $E$ and $\fem$ on an event-by-event basis.  This method has
been published elsewhere\cite{groom07,RPP12}, but the identity of the approach 
with the ``$Q/S$ method'' was not previously recognized.  {\color{blue}In the 2nd lines above, 
the expressions are rearranged
to make the required symmetry between $Q$ and $S$ more evident.}

\section{Discussion}\label{sec:discuss}

\begin{figure}
\centerline{\includegraphics[height=3in]{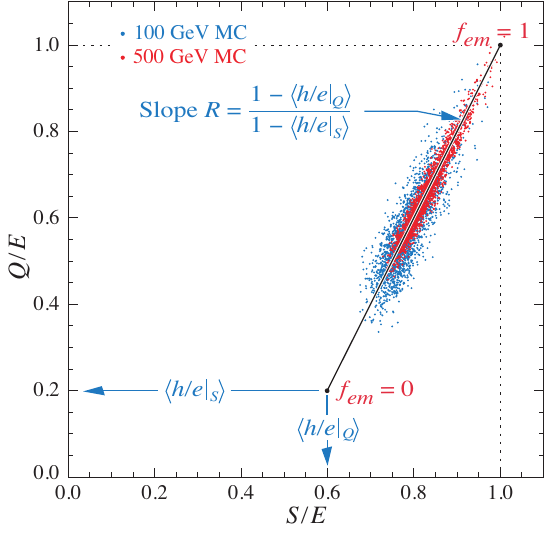}}
\caption{Energy-independent event locus in the $Q/E$--$S/E$ plane. With increased energy, resolution improves and the mean moves upward along the locus.}
\label{fig:locus}
\end{figure}

In part because of the relatively small number of particles involved early in a hadronic cascade,
the efficiency with which the hadronic energy deposit is visible in either the Cherenkov or scintillator channel varies from event to event. In contrast, the efficiency with which the EM deposit is detected varies little. The result is that $\eta_Q$ and $\eta_S$ are stochastic variables, mostly reflecting the variation of $h$.  The values of $\eta_Q$ and $\eta_S$ required to compute the energy estimator for each event via Eq.~\ref{eqn:EviaQoverS} (or Eq.~\ref{eqn:Elongform}) 
are not only unknown but unknowable, given ``only'' dual readout.  
In actual data reduction, there is little choice but to replace them by their mean values:
\begin{eqnarray}
E&=&\displaystyle 
\frac {S(1- \vev{\eta_Q}) - Q{\color{red}(1-}\vev{\eta_S}{\color{red})}}{\vev{\eta_S} -  \vev{\eta_{\color{red}Q}}}\\
\color{blue}\fem&\color{blue}=&\color{blue} \displaystyle\frac      {Q\vev{\eta_S}-S\vev{\eta_Q}}     {S(1-\vev{\eta_Q}) -Q(1-   \vev{\eta_S})  }
\label{eqn:aveElongform}  
\end{eqnarray}

It is also useful to rewrite Eqs.~\ref{eqn:Q} and \ref{eqn:S}:
\begin{eqnarray} 
\vev{Q/E} &=&  f_{em} + (1-f_{em}) \, \vev{\eta_Q}
\label{eqn:meanQoE}\\
\vev{S/E} &=& f_{em}+ (1- f_{em})  \, \vev{\eta_S}
\label{eqn:meanSoE}
\end{eqnarray}
Since $\vev{Q/E}$ and $\vev{S/E}$ are linear in $\fem$,  $\vev{Q/E}$ is a linear function 
of $\vev{S/E}$, describing a line segment from $(\vev{Q/E} , \vev{S/E}) = (\vev{\eta_Q}, \vev{\eta_S})$ at the all-hadronic extreme, $\fem=0$, to $(\vev{Q/E} , \vev{S/E}) = (1,1)$, 
at the all-EM extreme, $\fem=1$.  This event locus is shown in Fig.~\ref{fig:locus}.  As the energy increases, the Monte Carlo event scatter shown in the figure moves upward and becomes more
clustered as the resolution improves.

The energy-independent event locus has slope
\begin{equation}
R = \displaystyle \frac {1- \vev{\eta_Q}}{1- \vev{\eta_S }} \ .
\label{eqn:slopedefinition}
\end{equation}
This slope can be determined either by linear fits to monoenergetic (test beam) event distributions in the $Q/E$--$S/E$ plane, or, perhaps more accurately, 
by separately finding $\vev{\eta_Q}$ and $\vev{\eta_S}$ via $\pi/e$ measurements as a function of energy.  It can be used to cast Eq.~\ref{eqn:Elongform} into a more tractable 
form\cite{groom07,RPP12}:
\begin{equation}
E = \frac{RS-Q}{R-1}
\label{eqn:Eshortform}
\end{equation}

Since $\fem$ is not needed in data reduction, it is only of academic interest. Experimental distributions based on DREAM data are shown in Refs.~\cite{Wigmans_Perugia_04} and \cite{DREAM05} (Ak05). These are broadened by resolution effects, and so do not necessarily conform to $0\leq\fem\leq1$.  

\section*{Acknowledgments}

John Hauptman's critical comments and suggestions have been particularly helpful.
This work was supported by the U.S. Department of
Energy under Contract No.\ DE-AC02-05CH11231.

\end{document}